\documentclass[preprint]{aastex}
\usepackage{amsbsy} 

\slugcomment{submitted to MNRAS}

\shorttitle{Externally Fed Accretion onto Protostars}
\shortauthors{Dalba \& Stahler}

\begin{document}

\title{Externally Fed Accretion onto Protostars}

\author{Paul A. Dalba\altaffilmark{1} and Steven W. Stahler\altaffilmark{1}}

\altaffiltext{1}{Astronomy Department. University of California,
Berkeley, CA 94720}

\email{Sstahler@astro.berkeley.edu}

\begin{abstract}
The asymmetric molecular emission lines from dense cores reveal slow, inward
motion in the clouds' outer regions. This motion is present both before and 
after the formation of a central star. Motivated by these observations, we 
revisit the classic problem of steady, spherical accretion of gas onto a 
gravitating point mass, but now include self-gravity of the gas and impose a 
finite, subsonic velocity as the outer boundary condition. We find that the 
accretion rate onto the protostar is lower than values obtained for isolated, 
collapsing clouds, by a factor that is the Mach number of the outer flow. 
Moreover, the region of infall surrounding the protostar spreads out more 
slowly, at a speed close to the subsonic, incoming velocity. Our calculation, 
while highly idealized, provides insight into two longstanding problems -- the 
surprisingly low accretion luminosities of even the most deeply embedded 
stellar sources, and the failure so far to detect spatially extended, 
supersonic infall within their parent dense cores. Indeed, the observed 
subsonic contraction in the outer regions of dense cores following star 
formation appears to rule out a purely hydrodynamic origin for these clouds.
\end{abstract}


\section{Introduction}

Astronomers are elucidating, in much greater detail than ever before, the 
earliest phases of low-mass star formation. It has long been known that at 
least half of dense cores within molecular clouds do not yet contain young 
stars \citep{b86}. It is also known that these starless cores have, on average,
less mass than cores with embedded stars \citep{jma99}. What is now clearer is 
the chemical and structural progression of cores as they approach the collapse
threshold and then cross it. The most centrally concentrated starless cores 
tend to exhibit the asymmetric emission line profiles signifying inward 
contraction, with the inferred velocity being several tenths of the sound speed 
\citep{lmt01}. The same blueward asymmetry is seen in those cores with 
embedded stars, where it is especially common \citep{m97}. Detailed examination
of line profiles reveals no qualitative change across the collapse threshold, 
except for the appearance of broad wings that are attributable to molecular 
outflows \citep{ge00}.

It is likely that the contraction observed in starless cores is driven by
self-gravity, as material in the core's outer region settles toward the center
\citep[e.g.,][]{cb01}. A similar, exterior contraction should also occur in
cores that have produced stars, and observations thus far support this 
conclusion. Here, targeted studies are rare, in part because outflows often 
contaminate the emission line spectra. As one example, \citet{m96} utilized
a simple radiative transfer model to measure the spatially averaged contraction
speed in L1526, a core with an embedded star that drives a weak outflow. Their 
derived contraction speed of 0.025~km~s$^{-1}$ is indeed close to the inferred 
speeds in starless cores. The CS and H$_2$CO spectra for the low-mass infall 
candidates identified by \citet{m97} have a similar degree of asymmetry as 
L1526, but have not been modeled to this level of detail.  

Taken as a whole, these observations suggest that a dense core reaches the 
point of collapse by accruing matter externally. The core itself is a 
relatively quiescent structure, since its observed emission lines have nearly 
thermal widths \citep{bg98}. The exterior molecular gas, however, has typical 
speeds of 1 to 2~km~s$^{-1}$, corresponding to Mach numbers of 5 to 
10 \citep{bt07}. How exactly material passes from this environment into the 
core is unknown, despite increasingly detailed mapping of the transition 
region \citep{p10}. In any event, the core is dynamically evolving while it 
collects mass. Gas in its outer mantle creeps inward, and continues to do so 
even after the central region collapses to form a star. 

The standard theory of collapse, first formulated prior to the discovery of 
dense cores, paints a very different picture. In the still widely used 
self-similar model of \citet{s77}, the cloud is a perfectly static entity 
before the first appearance of a protostar. The impulsive suctioning of gas 
caused by this event sends out a sharp rarefaction wave, expanding at the speed
of sound. It is within this wavefront that infall occurs. Gas leaves the
rarefaction wave with zero velocity and attains the sound speed roughly halfway
inside it.

For a typical dense core radius of 0.1~pc and sound speed of 0.2~km~s$^{-1}$, 
the rarefaction wave reaches the boundary in 0.5~Myr. This interval, or
one comparable, has been traditionally considered to be the lifetime of 
the protostar phase. Comparing the dynamical picture from theory with 
observations, it is striking that there are no known cores with embedded stars 
in which the sonic transition occurs reasonably close to the cloud boundary. On
the contrary, the inferred location of this point in the best spectroscopic
collapse candidates is only 0.01-0.02~pc from the protostar 
itself \citep{c95,ge97,df01,b02}. Again, we must remember that strong molecular 
outflows can both mask and mimic large-scale motion in these studies. On the 
other hand, submillimeter continuum mapping of cores with embedded stars also 
indicates that infall is confined to a small, central region. \citet{ser02} 
have shown that more widespread collapse would give rise to a change in the 
interior density profile that is not observed. The continuing absence of dense 
cores that contain stars and exhibit true global infall is forcing us to 
conclude either that all observed protostars happen to be especially young, or 
else that infall actually spreads more slowly than traditional theory posits.

The collapse model of \citet{s77} also predicts that the mass accretion rate
onto the protostar is a constant in time:
\begin{equation}
{\dot M} \,=\, m_\circ\,{c_s^3 \over G} \,\,.
\end{equation}
Here, $c_s$ is the isothermal sound speed in the core and $m_\circ$ is a 
nondimensional number with a value of 0.975. Many authors, starting with
\citet{k90}, have noted that the accretion luminosity associated with this rate
exceeds those observed for embedded, infrared sources. Subsequent numerical 
simulations relaxed the assumption of self-similarity, but usually imposed a 
rigid boundary on the dense core, i.e., a fixed surface where the fluid 
velocity vanishes \citep[e.g.,][]{fc93,o99,vb05}. In these calculations, 
${\dot M}$ varies in time, but is even higher, at least before gas begins to 
drain from the boundary region. The sonic transition occurs only slightly 
inside this boundary \citep[see Fig.~1 of][]{fc93}.

In response to the persistent and growing disparity between theory and 
observation, researchers have taken two very different paths. Some have focused
on the possibility that, while cloud collapse occurs in the traditional manner,
infalling gas is retained in a circumstellar disk, and only released 
periodically in bursts onto the protostar \citep[e.g.,][]{dv12}. If the 
duration of bursts is sufficiently brief, we could be viewing protostars only 
when $\dot M$ is relatively low. Whatever the merits of this idea, it does not 
address the troubling aspects of collapse models generally. Protostellar 
luminosity and cloud collapse are clearly linked, and a solution that considers
both is preferable. 

Following this second, unified path, \citet{f04} generalized the model 
of \citet{s77} and investigated the self-similar evolution of a cloud that has 
a specified velocity profile at the start. \citet{go07} and \citet{go09} 
simulated numerically the birth of a dense core as arising from the convergence
of a supersonic inflow. Both groups tracked the evolution past the start of core
collapse. \citet{sy09} used perturbation theory to find the contraction in 
starless cores that are just approaching collapse, and were able to reproduce 
the asymmetric emission lines \citep{sy10}. 

In this brief contribution, we focus on how continuous, external mass addition 
to the dense core affects protostellar infall itself. We show that $\dot M$ in 
this case is equal to $c_s^3/G$ times a factor that is essentially the Mach 
number of the incoming flow. This key result is easy to understand physically. 
For a cloud that is marginally stable against self-gravity, the free-fall and 
sound-crossing times are comparable. It follows that the mean speed of 
infalling fluid elements in an isolated cloud undergoing collapse is 
approximately $c_s$. When accretion is imposed from the outside, this mean 
speed is closer to the externally impressed value. The mass transport rate is 
lowered by their ratio, which is the incoming Mach number.

Since the contraction speed in the outer portion of the dense core is well 
under $c_s$, the reduction in $\dot M$ may be substantial. The accretion 
luminosity is similarly reduced from traditional values, alleviating the 
discrepancy between theory and observation in this regard. We further show that
the region of infall spreads out at a speed that is close to the subsonic 
injection velocity in the core's exterior. This relatively slow expansion helps
to explain why global, supersonic infall has been difficult to observe.

In Section~2 below, we describe the physical problem to be solved, present the
relevant equations in nondimensional form, and then give our method of solution.
Section~3 presents numerical results. Finally, Section~4 compares our findings
with those of other studies, and indicates future directions for extending this 
work.

\section{Formulation of the Problem}
\subsection{Physical Assumptions}

Dense cores are supported against self-gravity not only by thermal pressure,
but also by the interstellar magnetic field, which penetrates their interior
\citep{c99}. The limited data from polarization mapping indicate that the field
has a relatively smooth spatial variation \citep{rgm11}, as would be expected 
from the absence of significant MHD wave motion. From a theoretical 
perspective, the objects are essentially magnetostatic structures. They must 
evolve quasi-statically, however, since at some point they undergo gravitational
collapse. This evolution is facilitated both by ambipolar diffusion, i.e.,
the slippage of neutral gas across field lines, and by the addition of fresh
material from the core's surroundings.

There is no reason why the dense core's accrual of gas should cease once that
object has produced a central star. Prior to this event, the core had an 
extended period of slow evolution. Thus, the {\it internal} accretion rate onto
the protostar has had ample time to adjust so as to match the {\it external} 
rate at which mass is continually added to the dense core. In other words, 
$\dot M$ is determined by conditions far from the star itself. This was also 
the case in the traditional picture of an isolated cloud undergoing collapse, 
where $\dot M$ is set by the global free-fall time. A new picture, more 
consistent with observations, is that a dense core harboring a stellar source 
is a temporary repository for material that drifts in from the environment on 
its way to the protostar.\footnote{The near equality of internal and external 
accretion rates does {\it not} hold if the external gas is injected 
supersonically. In that case, simulations find that the inner rate undergoes a 
sharp, transient burst just after star formation, before ultimately 
equilibrating to the outer rate \citep[see Fig.~13 of][]{go09}.} 

To begin exploring the new picture qualitatively, we adopt a very simple model.
We ignore magnetic forces entirely, and treat the dense core as a region of
steady, isothermal flow surrounding the protostar. Recent observations by the
Herschel satellite have shown vividly how dense cores are nested inside
pc-scale filaments \citep{m10}. Hence, the inward flow of gas is assuredly
anisotropic. We ignore this complication as well, and assume spherical 
symmetry. Every fluid element experiences both the gravitational force from the
protostar and that from interior cloud gas. Our problem is thus identical to 
the classic inflow calculation of \citet{b52}, except that we include 
self-gravity of the gas.

Like Bondi, we apply the inner boundary condition that this material is in 
free fall onto the star. However, we no longer demand that the density approach
a fixed value at spatial infinity \citep[as did][]{c78}. Instead, we require 
that the {\it velocity} reach some imposed, subsonic value. Note again the 
critical difference from the many numerical collapse simulations, where the 
infall velocity was set to zero at some radius, effectively isolating the dense
core from its surroundings. In effect, our imposition of a finite, bounding 
velocity means that we are treating protostellar accretion as a flow problem, 
rather than a collapse, which would necessitate a time-dependent treatment 
\citep[see the discussion in][]{w84}.

In reality, of course, the subsonic flow begins at a finite boundary. By
taking this boundary to infinity, we are focusing, again for simplicity, on the
earliest phase of protostellar accretion, when the mass of the central object 
is much less than that of the parent dense core.\footnote{Specifically, we are 
requiring that \hbox{$M_\ast\,\ll\,r_b\,c_s^2/G$}, where $r_b$ is the dense 
core boundary. The dense core itself is marginally unstable gravitationally, 
so the righthand side of this inequality is close to the core mass.} Since 
observations indicate that the {\it final} protostar mass is no more than 
20 to 30~percent that of the core \citep{all07,r09}, this early epoch spans a 
substantial fraction of the protostar's total lifetime. 

\subsection{Quasi-Steady Flow}

In spherical symmetry, momentum conservation in an isothermal, steady-state 
flow reads
\begin{equation}
u\,{{du}\over{dr}} \,=\, -{{G \left(M_\ast + M_r\right)}\over r^2} \,-\,
{c_s^2\over\rho}\,{{d\rho}\over{dr}} \,\,.
\end{equation}
Here, $u$ is the fluid velocity, taken to be positive for inward motion. The
quantities $M_\ast$ and $M_r$ are, respectively, the mass of the central
protostar and that of the cloud gas interior to radius $r$. Finally, $\rho$ is
the mass density of the gas and $c_s$ the isothermal sound speed. The cloud
mass $M_r$ and density $\rho$ are related through
\begin{equation}
      {{d M_r}\over{dr}} \,=\, 4\,\pi\,r^2\,\rho \,\,.
\end{equation}
Since the flow is steady, the same mass per unit time crosses every spherical
shell. Denoting this mass transport rate by $\dot M$, we have
\begin{equation}
      {\dot M} \,=\, 4\,\pi\,r^2\,\rho\,u \,\,.
\end{equation}

The inner boundary condition is that the velocity approach free fall onto the 
star:
\begin{equation}
 \lim_{r \rightarrow 0} \,u \, =\, \sqrt{{2\,G\,M_\ast}\over r} \,\,.
\end{equation} 
Equivalently, $M_r$ must approach zero in this limit. We set our outer boundary
condition at spatial infinity and demand that
\begin{equation}
\lim_{r \rightarrow \infty} u\,(r) \,=\, u_\infty \,\,,
\end{equation}
where $u_\infty$ is some fixed, subsonic velocity that is consistent with the
spectroscopic observations. 

Equation~(4) incorporates our assumption that the inner accretion rate onto the
star matches the outer rate imposed externally. In more detail, the match will
not be exact, and the difference drives secular changes in the gas density
and total core mass that only a full evolutionary calculation can track. Within
the infall region, the assumption of steady state is justified as long as the 
crossing time through that region is brief compared to the time for $M_\ast$ to
increase significantly. After completing the calculation, we will verify its 
self-consistency in this regard in Section~4 below. 

\subsection{Nondimensionalization and Method of Solution}

It is convenient to recast our dynamical equations into nondimensional form. We
adopt, as basic quantities, $G$, $M_\ast$, and $c_S$. The solution is then
characterized by a dimensionless parameter, the Mach number associated with the 
inflow at the boundary:
\begin{equation}
\beta \,\equiv\, u_\infty/c_s \,\,.
\end{equation}
We define a nondimensional velocity as \hbox{${\tilde u}\,\equiv\,u/c_s$}, and 
a nondimensional cloud mass as \hbox{${\tilde M_r}\,\equiv\,M_r/M_\ast$}. In 
addition, we define a fiducial radius, density, and mass transport rate,
respectively, as 
\begin{eqnarray}
r_\circ &\,\equiv\,& {{G\,M_\ast}\over{c_s^2}} \,\,,\\
\rho_\circ &\,\equiv\,& {c_s^6 \over{G^3\,M_\ast^2}} \,\,,\\
{\dot M}_\circ &\,\equiv\,& {c_s^3\over G} \,\,,
\end{eqnarray}
and set \hbox{${\tilde r}\,\equiv\,r/r_\circ$} and 
\hbox{${\tilde \rho}\,\equiv\,\rho/\rho_\circ$}.

Our dynamical equations become
\begin{eqnarray}
{\tilde u}\,{{d{\tilde u}}\over{d{\tilde r}}} &\,=\,& 
-{{1 + {\tilde M}_r}\over{\tilde r}^2} \,-\,
{1\over{\tilde\rho}}\,{{d{\tilde\rho}}\over{d{\tilde r}}} \,\,, \\
{{d{\tilde M}_r}\over{d{\tilde r}}} &\,=\,& 4\,\pi\,{\tilde r}^2\,
{\tilde \rho} \,\,,\\
 \lambda \,&=&\, 4\,\pi\,{\tilde r}^2\,{\tilde\rho}\,{\tilde u} \,\,,
\end{eqnarray}
where \hbox{$\lambda\,\equiv\,{\dot M}/{\dot M}_\circ$}. The boundary
conditions are
\begin{eqnarray}
\lim_{{\tilde r} \rightarrow 0} \,{\tilde u} &\,=\,&
\sqrt{2\over{\tilde r}} \,\,, \\
\lim_{{\tilde r} \rightarrow \infty} {\tilde u} &\,=\,& \beta \,\,.
\end{eqnarray}
From now on, we drop all tildes and alert the reader whenever we revert to
dimensional variables.

To solve this system, we first combine equations~(12) and (13) into
\begin{equation}
{{d M_r}\over{dr}} \,=\, {\lambda\over u} \,\,.
\end{equation}
Next, we take the logarithmic derivative of equation~(13) and use the result to
eliminate the density gradient in equation~(11):
\begin{equation}
\left(u\,-\,{1\over u}\right) {{du}\over{dr}} \,=\, 
-{{1 \,+\,M_r} \over r^2} \,+\, {2\over r} \,\,.
\end{equation}
The basic strategy is to solve equations~(16) and (17) to obtain $u (r)$ and
$M_r (r)$. Equation~(13) then provides an algebraic relation for $\rho (r)$. 

Before describing the solution technique in more detail, we first derive a 
simple relation between $\lambda$ and $\beta$. First note that equation~(16)
tells us that $dM_r/dr$ asymptotes to a finite value for large $r$. Thus, 
$M_r$ itself increases without limit. On the other hand, since $u$ approaches a
finite value, $du/dr$ must vanish asymptotically. The entire lefthand side of 
equation~(17) thus also goes to zero. The vanishing of the righthand side of 
this equation leads us to conclude that
\begin{equation}
\lim_{r \rightarrow \infty} M_r \,=\,2\,r \,\,,
\end{equation}
confirming our previous claim that $M_r$ increases indefinitely.

Equation~(18) implies that
\begin{equation}
\lim_{r \rightarrow \infty} {{dM_r}\over{dr}} \,=\,2 \,\,.
\end{equation}
We now apply this same, large-$r$ limit to equation~(16). After
utilizing equation~(15), we find the desired relation:
\begin{equation}
\lambda \,=\, 2\,\beta \,\,.
\end{equation}
In dimensional language, this is
\begin{equation}
{\dot M} \,=\, \left({{2\,u_\infty}\over c_s}\right){{c_s^3}\over G} \,\,.
\end{equation}
Here we see explicitly the reduction factor modifying the conventional mass
accretion rate. 

Supplied with this result, we proceed to solve the coupled equations~(16) and
(17) as follows. We start at an $r$-value inside the sonic point $r_s$, i.e., 
the radius where \hbox{$u\,=\,1$}. At this interior point $r_i$, we guess both
$u$ and $M_r$. Integrating outward, we find that $u (r)$ diverges upward or 
downward as we approach $r_s$, depending on the guessed $u (r_i)$. Using a 
bifurcation technique, we find that value of $u (r_i)$ which extends the 
solution farthest out. We then integrate inward from $r_i$, and similarly find 
that the velocity either climbs above or falls below the free-fall value given 
by equation~(14). A similar bifurcation technique then yields the best-fit 
$M_r$ at \hbox{$r\,=\,r_i$}.
 
Having satisfied the inner boundary condition and established the location of
the sonic point $r_s$, our next task is to cross this point. We apply
L'H$\hat{\rm {o}}$pital's rule to equation~(17) and use equation~(16) to find
\begin{equation}
{{du}\over{dr}} \,=\, -{\sqrt{1\,-\,\lambda/2} \over r} \,\,,
\qquad{\rm at}\,\,\, r\,=\,r_s \,\,.
\end{equation}
This relation, together with the numerically established $M_r (r_s)$, allows us
to integrate outward. As long as $\lambda$ obeys equation~(20), the velocity
$u$ automatically approaches $\beta$ at large $r$.

\section{Numerical Results}

Figure~1 displays the velocity profile $u (r)$ obtained in the manner 
indicated. Here we have set the incoming Mach number equal to the 
representative value \hbox{$\beta \,=\, 0.2$}. Close to the origin, the velocity
diverges as gas freely falls onto the protostar. Farther out, $u$ first dips
below $\beta$ and then asymptotically approaches it through a series of 
decaying oscillations.

The sonic point is found to be \hbox{$r_s\,=\,0.544$}, a slight increase over
the Bondi value of 0.5. The {\it dimensional} sonic radius is then
\begin{mathletters}
\begin{eqnarray}
r_s &\,=\,& \lambda\,{\tilde r_s}\,c_s\,t \\
{\phantom u} &\,=\,& 2\,{\tilde r_s}\,\beta\,c_s\,t \,\,. 
\end{eqnarray}
\end{mathletters}
Thus, the point moves out with speed
\begin{equation}
{{d r_s}\over{dt}} \,=\, 2\,{\tilde r_s}\,\beta\,c_s \,\,.
\end{equation}
If we identify the infall region as that volume interior to $r_s$, then this
region expands at \hbox{$0.22\,c_s$}. Note again the contrast with the model of
\citet{s77}, where the equivalent speed is exactly $c_s$. Note also that the 
value of $2\,{\tilde r_s}$ is close to unity and insensitive to $\beta$, rising 
only to 1.22 for \hbox{$\beta\,=\,0.4$}. In summary, the infall region spreads 
out at a speed close to $\beta\,c_s$, at least during this early epoch.

Figure~2 shows the density profile $\rho (r)$. Inside $r_s$, the profile is 
that associated with freely falling gas. From equations~(13), (14), and (20), 
this limiting profile is
\begin{equation}
\rho \,=\,{\beta\over
{2\,\sqrt{2}\,\pi\,r^{3/2}}} \,\,\qquad\,\, r\,\ll\,r_s \,\,.
\end{equation}
The dimensional equivalent is
\begin{equation}
\rho \,=\,\left({\beta\over{c_s\,t}}\right)^{1/2}
\!{{c_s^2}\over{4\,\pi\,G\,r^{3/2}}}  \,\,,
\end{equation}
where we have also used \hbox{$M_\ast\,=\,{\dot M}\,t$}. At early times, the 
density declines throughout the free-fall region, while the region itself 
expands.

Outside of $r_s$, equations~(13), (15) and (20) tell us that
\begin{equation}
\rho\,=\,{1\over{2\,\pi\,r^2}} \,\,\qquad\,\, r\,\gg\,r_s \,\,.
\end{equation}
which is dimensionally equivalent to
\begin{equation}
\rho  \,=\, {c_s^2\over{2\,\pi\,G\,r^2}} \,\,.
\end{equation}
This is the profile of the singular isothermal sphere, as we expect for a
self-gravitating gas cloud that contains a star of relatively small mass and is
close to hydrostatic equilibrium \citep[][Chapter~9]{sp04}.

\section{Discussion}
Having solved the problem as posed, our next task is to check that the 
assumption of steady-state flow is reasonable. The outer portion of the cloud
evolves slowly with time, as a structure that is never far from hydrostatic
balance. It is the interior region that concerns us. Again, we need to make 
sure that the crossing time from $r_s$ to the origin is less than
$M_\ast/{\dot M}$, which in turn is close to the evolutionary time 
$t$.\footnote{The time $M_\ast/{\dot M}$ would {\it not} be close to $t$ if
$\dot M$ varied rapidly in time, as it does in simulations of isolated,
collapsing clouds. However, we do not expect such rapid variation when a
subsonic, external flow is imposed from the start.} Thus, we require
\begin{equation}
\int_0^{r_s} \!{{dr}\over u} \,<\,t \,\,.
\end{equation}
But \hbox{$u\,>\,c_s$} in this region, so that the integral is less than
$r_s/c_s$. Since the infall region expands subsonically, this ratio is indeed
less than $t$.

We stress the need for a fully time-dependent calculation to verify our
main results. Assuming there is no qualitative change, it is instructive to
compare our findings with those of \citet{go09}. Using direct simulation, these
authors built up a dense core from a spherically converging, supersonic flow. 
The accretion shock bounding the core first expands and then contracts 
supersonically before the core itself undergoes interior collapse. When the 
protostar first appears, the velocity profile throughout the dense core is 
uniform and supersonic, with \hbox{$u\,\approx\,3\,c_s$}. A rarefaction wave 
erupts from the origin, but now expands supersonically. The wave quickly 
overtakes the accretion shock and effectively destroys it, so that the 
converging inflow thereafter directly impacts the protostar. \citet{go11} 
repeated the simulation using a planar, external flow and found substantially 
the same results.

We observe once more the close relationship between the velocity of $r_s$ and 
that of the gas introduced externally. To see this connection in yet another 
light, note that $r_s$ is located roughly where the free-fall velocity onto the
star matches the sound speed:
\begin{equation}
\sqrt{{2\,G\,M_\ast}\over{r_s}} \,\approx \, c_s \,\,.
\end{equation}
If we let \hbox{$M_\ast\,=\,{\dot M}\,t$} and use equation~(21), we find that
\hbox{$r_s\,\approx\,u_\infty\,t$}.

The fact that the outer regions of dense cores are contracting subsonically 
{\it both before and after} the appearance of a protostar appears to rule out a 
purely hydrodynamic origin for these objects. The many researchers who have
simulated cluster formation in turbulent molecular gas have effectively 
sidestepped this issue. In these studies, random velocity fluctuations are
applied to a representative cube of molecular gas. Once these fluctations 
thoroughly stir up the gas, self-gravity is switched on and overdense structures
promptly collapse \citep[see, e.g.,][]{bhv99,hmk01,pn02,okm08}.  

More realistically, a dense core is self-gravitating and gains mass prior to 
this collapse, and a force in addition to the thermal pressure gradient must 
support it during this early epoch. This force is already known -- the pressure
gradient associated with the interstellar magnetic field. The field may not 
play a key role in the collapse phase studied here, although it apparently 
impacts the formation of circumstellar disks \citep{lks11}. In any case, the 
magnetic field is surely key in the early buildup of dense cores within their 
parent cloud filaments, and must also mediate the transition between the 
supersonic exterior of each core and its supersonic exterior.

\acknowledgments
We are grateful to Paola Caselli, Neal Evans, and Phil Myers for sharing their 
expertise on dense cores and the subtle ways in which observations are 
revealing their dynamics. SWS was partially supported by NSF Grant AST-0908573.

\clearpage

\begin{figure}
\plotone{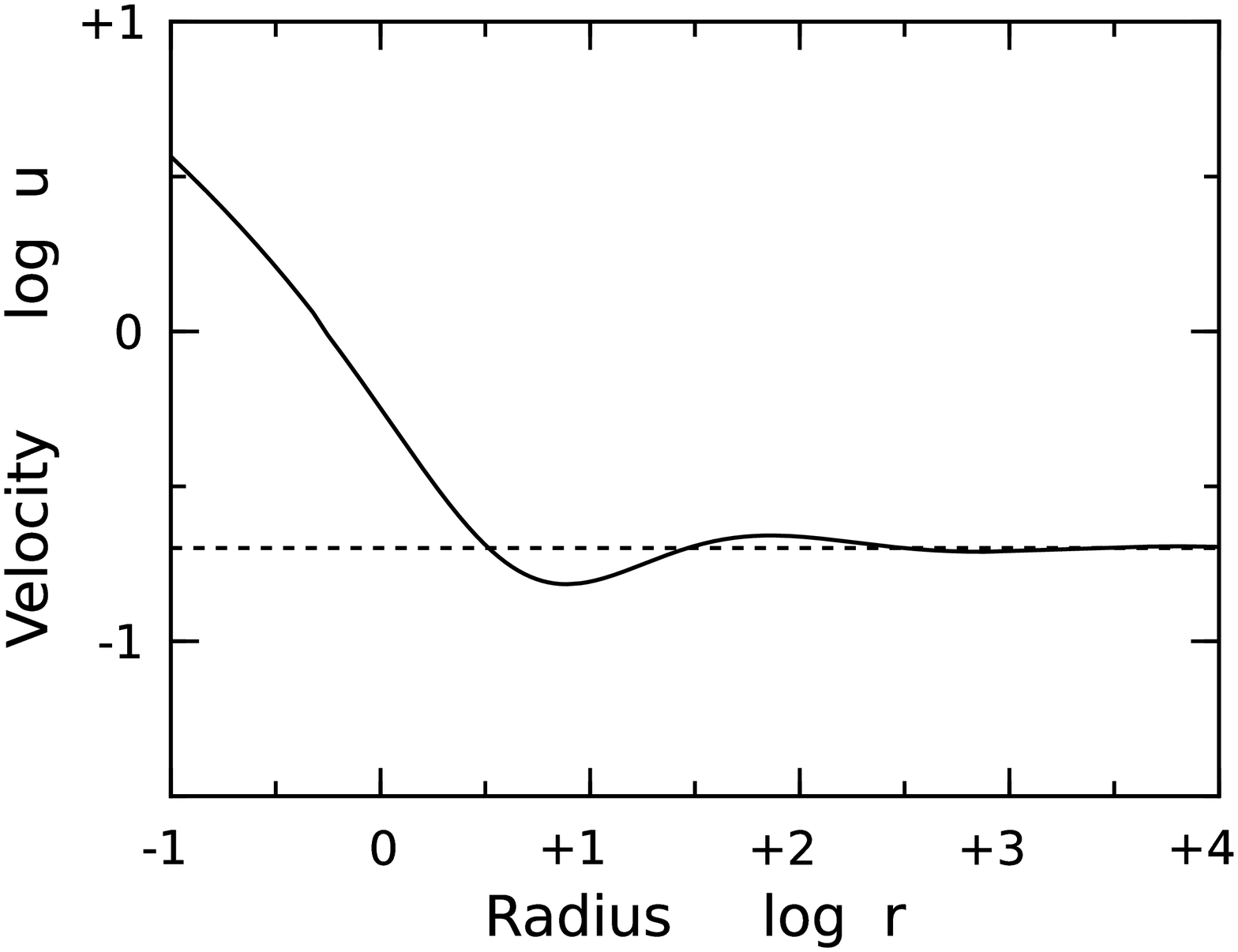}
\caption{Velocity profile in the quasi-steady flow, for the representative case
\hbox{$\beta\,=\,0.2$}. Both the velocity and radius are in the nondimensional
units defined in the text. The velocity asymptotically approaches the imposed,
outer value, shown as the dashed, horizontal line.}
\end{figure}

\begin{figure}
\plotone{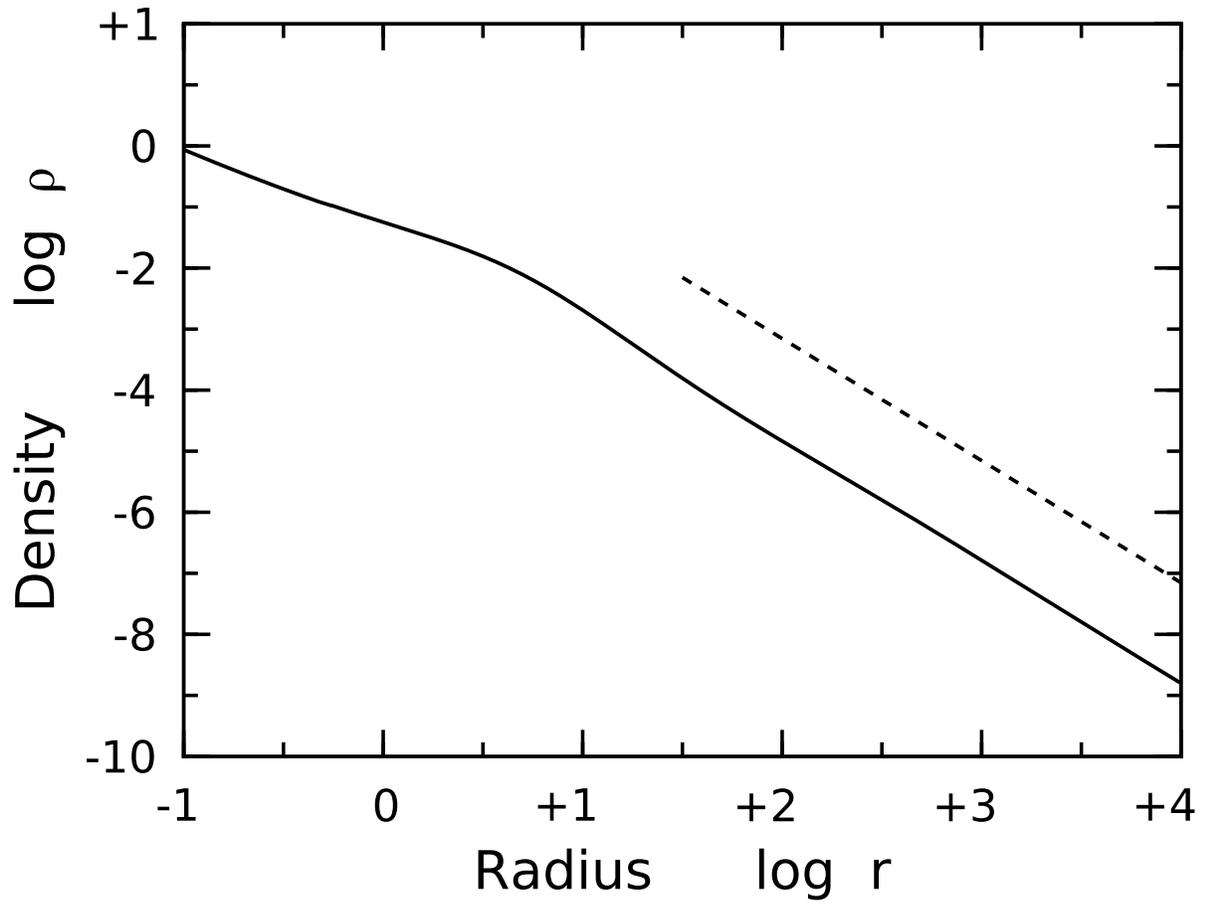}
\caption{Density profile in the quasi-steady flow, for \hbox{$\beta\,=\,0.2$}.
The density and radius are both nondimensional. The dashed curve is a density
profile that varies as $r^{-2}$.}
\end{figure}

\end{document}